\documentclass[12pt]{article}

\usepackage[mathscr]{eucal}
\usepackage[utf8]{inputenc}
\usepackage{slashed}
\usepackage{subcaption}
\usepackage{feynmp}
\usepackage{tikz-feynman}
\tikzfeynmanset{compat=1.1.0}
\usepackage{slashed}
\usepackage{amssymb,amsmath}

\begin{document}

\begin{titlepage}

\begin{flushright}
arXiv:2304.03025
\end{flushright}

\vspace{1ex}

\begin{center}
{\Large \bf Perturbative Aspects of CPT-Even Lorentz-\\
Violating Scalar Chromodynamics}
\end{center}

\begin{center}
{\large B. Altschul,$^{1}$\footnote{{\tt altschul@mailbox.sc.edu}}
L. C. T. Brito,$^{2}$\footnote{{\tt lcbrito@ufla.br}}
J. C. C. Felipe,$^{3}$\footnote{{\tt jean.cfelipe@ufvjm.edu.br}}
S. Karki,$^{1}$\footnote{{\tt karkis@email.sc.edu}}\\
A. C. Lehum,$^{4}$\footnote{{\tt lehum@ufpa.br}}
A. Yu. Petrov,$^{5}$\footnote{{\tt petrov@fisica.ufpb.br}}}

\vspace{5mm}
$^{1}${\sl Department of Physics and Astronomy,} \\
{\sl University of South Carolina, Columbia, SC 29208} \\
$^{2}${\sl Departamento de F\'{i}sica, Instituto de Ci\^{e}ncias Naturais} \\
{\sl Universidade Federal de Lavras, Caixa Postal 3037, 37200-900, Lavras, MG, Brasil} \\
$^{3}${\sl Instituto de Engenharia, Ci\^{e}ncia e Tecnologia,} \\
{\sl Universidade Federal dos Vales do Jequitinhonha e Mucuri, Avenida Um} \\
{\sl 4050, 39447-790, Cidade Universit\'{a}ria, Jana\'{u}ba, MG, Brazil} \\
$^{4}${\sl Faculdade de F\'{i}sica, Universidade Federal do Par\'{a}, 66075-110, Bel\'{e}m, PA, Brazil} \\
$^{5}${\sl Departamento de F\'{i}sica, Universidade Federal da Para\'{i}ba,} \\
{\sl Caixa Postal 5008,	58051-970 Jo\~{a}o Pessoa, PB, Brazil}

\end{center}

\medskip

\centerline {\bf Abstract}

\bigskip

In this work, we formulate the theory of Lorentz-violating scalar Quantum Chromodynamics
with an arbitrary non-Abelian gauge group. This theory belongs to the class of models encompassed by the
standard model extension framework. At the lowest order in the theory's Lorentz violation parameters, we
calculate the divergent quantum corrections, including the renormalization group $\beta$-functions of the
theory. The Lorentz-violating sector is shown to be scale invariant if there is a particular relation between
the couplings.

\bigskip

\end{titlepage}

\newpage

\section{Introduction}

The physics of elementary particles rest on well-tested principles of symmetry. The presumptions 
that there are certain exact spacetime symmetries (described by the Lorentz group) and internal symmetries
(underlying strong and electroweak physics and described by an overall non-Abelian gauge group) 
in the current standard model (SM) are key examples. However, there is also a common understanding that the
SM as we observe it is really just an effective theory, describing low-energy elementary particle
interactions using a renormalizable quantum field theory. Thus, any symmetry that is apparent at observable
scales may actually be just a low-energy approximation, with those symmetries being violated at more
fundamental levels. One natural way to obtain an extension of the currently understood SM is, therefore,
by relaxing at least one of the fundamental symmetries imposed on the theory. In this paper, we are
specifically concerned with the case of explicit breaking of some of the spacetime symmetries---specifically
the isotropy and Lorentz boosts symmetries, which together generate the Lorentz group.

One of the most important directions in the study of Lorentz symmetry breaking consists of formulating and
studying the possible Lorentz-breaking extensions of various field theoretic models. The most
important advancement in this area was the formulation of the Lorentz-violating (LV) standard model extension
(SME)~\cite{ColKost1,ColKost2}. The SME is an effective theory framework in which additional operators are
added to the action of the SM; these operators are structurally similar to the usual SM operators, but unlike
the terms in the usual SM Lagrange density (which are taken to be scalars under proper, orthochronous
Lorentz transformations), the SME operators may have free Lorentz indices. Since the
foundational work near the end of the last century, a large number of studies of classical and quantum aspects
of various LV theories---most commonly LV extensions of spinor quantum electrodynamics (QED)---have been
completed. (See, for example~\cite{KosPic,ourrev}.) In addition, the SME approach has also been generalized
to the include the presence of classical gravitation~\cite{KosGra}. In this context, the perturbative studies
of LV non-Abelian gauge theories are extremely natural. 

There have already been some interesting results obtained using perturbative analyses of LV non-Abelian
gauge theories coupled to spinor matter---notably including SME generalizations of the SM's quantum
chromodynamics (QCD) sector with quarks and gluons: first, the one-loop renormalization of LV
non-Abelian gauge theories
with fermions completed
(including chiral fermions) in  Refs.~\cite{ref-collad-3,Colladay:2007aj,ref-collad-1}; second,
perturbative generation
of the non-Abelian generalization of the
Carroll-Field-Jackiw term~\cite{ourYMCS}; third, perturbative generation of a non-Abelian
aether-like term~\cite{NAaether}. However, coupling of non-Abelian LV theories to scalar matter has not been
explored at all, leaving this sector as a very natural area to study. Explicitly, our aim is to extend the
calculations of two-, three- and four-point correlation functions, previously obtained in LV scalar QED
in~\cite{BaetaScarpelli:2021dhz,our-paper}, to the non-Abelian case. That is, we will be formulating and
studying scalar QCD, and thereby obtaining the one-loop divergent quantum corrections for the theory,
from which its renormalization group behavior may be determined.
More specifically, in this paper we are focusing on the analysis of the gluon-scalar interaction,
since the gluon self-interaction
and gluon-ghost interaction were studied in Refs.~\cite{ref-collad-3,Colladay:2007aj}.   
Throughout this paper, we use standard particle
physics conventions [natural units $c=\hbar=1$, and $(+---)$ as the spacetime signature]. 

The structure of the paper is as follows. In section~\ref{sec01}, we introduce the action for
our theory, including gauge fixing and ghost contributions. In section~\ref{sec-scalar-loops},
we discuss the generation of the non-Abelian aether term. In section 4,
we obtain the scalar-vector vertex functions, and in section 5 we study the $\beta$-functions
that describe the resulting renormalization group (RG) behavior. Our conclusions are presented in section 6.
There are also two appendices. Appendix~\ref{appendixPV} collects the Passarino-Veltman basis integrals
used throughout our calculations, and in appendix~\ref{appendixGSE}, the calculation of the gluon self-energy
in the presence of scalar matter is presented in a little more detail.

\section{CPT-Even LV Scalar Chromodynamics}

\label{sec01}

Let us consider the non-Abelian generalization of the model studied
in~\cite{BaetaScarpelli:2021dhz,our-paper}, described by the Lagrange density
\begin{eqnarray}
\label{eq01}
\mathcal{L}&=& (D_{\mu}\phi_i)^\dagger\left(\eta^{\mu\nu}+c^{\mu\nu}\right)
D_{\nu}\phi_i - m^2\phi_i^\dagger\phi_i - \frac{\lambda}{4} (\phi_i^\dagger\phi_i)^2 \\
&&{}-\frac{1}{4}F^{\mu\nu}_{a}F_{a\mu\nu}+\frac{1}{4}\kappa^{\mu\nu\alpha\beta}F_{a\mu\nu}F_{a\alpha\beta}+
\mathcal{L}_{GF}+\mathcal{L}_{{\rm ghost}}, \nonumber
\end{eqnarray}
where $a$ is the non-Abelian gauge group index [we may
sometimes specialize to the gauge group $SU(N)$, or even to the physical $SU(3)$
of QCD, for definiteness]; the scalar fields $\phi_i$ are in the adjoint representation
(meaning the octet in QCD); $F_a^{\mu\nu}=\partial^{\mu}A_a^{\nu}-\partial^{\nu}A_a^{\mu}
+gf_{abc}A_b^{\mu}A_c^{\nu}$ is the gluon field strength;
$D^{\mu}=\partial^{\mu} - ieA_a^{\mu}T_a$ is the covariant derivative, with $T_a$ being the
generators $[T_{a}]_{bc}=if_{abc}$
of the gauge group in the adjoint; and $c^{\mu\nu}$ and $\kappa^{\mu\nu\alpha\beta}$ are
dimensionless constant tensors that describe the CPT-even but LV operators in the scalar and vector
sectors. All  the scalars have the same mass $m$, which we take to be real, so that
gauge symmetry is not spontaneously broken --- meaning that $m^2 > 0$. (Studies of theories with both
Lorentz symmetry breaking and spontaneous gauge symmetry breaking will be undertaken in the future.)
Note that a term like $F^{\mu\nu}_{a}F_{a\mu\nu}$ with a sum over group indices may also be written as
trace over $F^{\mu\nu}F_{\mu\nu}$, in terms of the group-valued field strengths $F^{\mu\nu}$.

Prior to the inclusion of the gauge-fixing and ghost terms, the Lagrange density (\ref{eq01}) contains
two tensors that describe the Lorentz-violating backgrounds through which the scalar and vector fields
propagate. However, the number of physically meaningful parameters is actually fewer than one might expect,
based just on counting the number of parameters in the SME tenors.
Physically observable quantities cannot actually depend on $c^{\mu\nu}$ without also
depending on $\kappa^{\mu\nu\alpha\beta}$, through the specific linear combination
$c^{\mu\nu}+\kappa_{\alpha}{}^{\mu\alpha\nu}$~\cite{ref-altschul14}. This quantity measures the mismatch
between the effective metrics appearing in the kinetic terms for different sectors of the theory. If 
$c^{\mu\nu}+\kappa_{\alpha}{}^{\mu\alpha\nu}=0$, then the whole theory is actually nothing more than
standard scalar QCD, written in skewed coordinates. Whenever possible, it is desirable to have the triviality
of the Lorentz violation in this case be evident in the description of the theory.

For simplicity --- especially for when we shall be looking at the RG
$\beta$-functions --- we shall assume that $c^{\mu\nu}$ takes an (aether-like) traceless ($c^{\mu}{}_{\mu}=0$)
form $c^{\mu\nu}={Q_1} u^\mu u^\nu$, dependent on a single preferred null vector
$u^\mu$ with $u^2=0$. The Lorentz violation coefficient in the gauge sector will also depend solely on
$u^{\mu}$, taking the form
\begin{equation}
\label{eq-kappa}
\kappa^{\mu\nu\alpha\beta} = \frac{{Q_2}}{{Q_1}}\left(c^{\mu\alpha}\eta^{\nu\beta}
-c^{\mu\beta}\eta^{\nu\alpha}+\eta^{\mu\alpha}c^{\nu\beta}-\eta^{\mu\beta}c^{\nu\alpha}\right)
\end{equation}
in terms of $c^{\mu\nu}$.
(In the limit of vanishing coupling, $g=0$, a $\kappa^{\mu\nu\alpha\beta}$ of this form is indicative of
birefringence-free gauge boson propagation.)
The case mentioned above --- in which the apparent Lorentz violation is actually fictitious --- corresponds to
$Q_{1}+2Q_{2}=0$. When this relation is satisfied, the theory is actually just Lorentz-invariant scalar
QCD, but expressed in a coordinate system in which the distance along the light-front axis
direction $u^{\mu}$ is measured on a different scale than distances along other four-vector directions.
General linear transformations of the global coordinates may be used to change $u^{\mu}$, $Q_{1}$, and $Q_{2}$,
but the quantity $Q_{1}+2Q_{2}$ remains invariant under such transformations, meaning that it may be measured
independently of the choice of coordinate system.

With the simplified SME tensors, the Lagrange density (\ref{eq01}) becomes
\begin{eqnarray}
\label{eq03}
\mathcal{L}&=& (D_{\mu}\phi_i)^\dagger\left(\eta^{\mu\nu}+{Q_1}u^{\mu}u^{\nu}\right)
D_{\nu}\phi_i - m^2\phi_i^\dagger\phi_i - \frac{\lambda}{4} (\phi_i^\dagger\phi_i)^2 \\
&&{}-\frac{1}{4}F_a^{\mu\nu}F_{a\mu\nu}+Q_2 u^{\mu}u^{\nu}F_{a\mu}{}^{\alpha}F_{a\nu\alpha}+\mathcal{L}_{GF}
+\mathcal{L}_{{\rm ghost}}.
\nonumber
\end{eqnarray}
We must further include in the Lagrange density (\ref{eq01}) or (\ref{eq03})
a gauge-fixing term, together with the corresponding Faddeev-Popov ghost contributions.
The gauge-fixing term $\mathcal{L}_{GF}$ is a further LV generalization of the usual Lorenz-like
gauge condition used for non-Abelian gauge theories
\begin{equation}
\mathcal{L}_{GF}=\frac{1}{2\xi}\left(\partial^{\mu}A_{a\mu}+\frac{1}{2}
\kappa_{G}^{\mu\nu}\partial_{\mu}A_{a\nu}\right)^{2}. 
\end{equation}
In general, there is freedom to choose any $\kappa_{G}^{\mu\nu}$; however, the goal of selecting this
generalized gauge-fixing term is to have the simplest possible propagator for the gauge sector, with
the same $\kappa^{\mu\nu\alpha\beta}$ in the physical and pure gauge components of the propagator tensor.
It is fairly clear that to make this possible, there must be a specific relationship between the
physical $\kappa^{\mu\nu\alpha\beta}$ and the $\kappa_{G}^{\mu\nu}$ assigned to the gauge-fixing term,
with $\kappa_{G}^{\mu\nu}=4\lambda Q_{2}u^{\mu}u^{\nu}$ for some constant $\lambda$.
If $Q_{1}+2Q_{2}=0$, so that the apparent Lorentz violation in (\ref{eq01}) is actually fictitious (being just
a coordinate artifact), having a gauge-fixing term of this form should make it possible to eliminate an
appropriately chosen $\kappa_{G}^{\mu\nu}$ tensor via the same coordinate redefinition that sets $Q_{1}=Q_{2}=0$.
 
The structure of the $\mathcal{L}_{GF}$ term necessitates that there should also be a ghost term, which
takes the form
\begin{equation}
\mathcal{L}_{{\rm ghost}}=-\bar{C}_{a}\left(\partial^{\mu}D_{ab\mu}+\frac{1}{2}
\kappa_{G}^{\mu\nu}\partial_{\mu}D_{ab\nu}\right)C_{b},
\end{equation}
where $C_{a}$ and $\bar{C}_{a}$ are the Faddeev-Popov ghost and anti-ghost field, respectively.
Once again, if the Lorentz violation is unphysical, we hope, with the right coordinate transformation,
to be able to eliminate the Lorentz violation in $\mathcal{L}_{{\rm ghost}}$ along with the other terms.

Since we shall be evaluating loop integrals only to first order in the Lorentz violation coefficients,
we may neglect anything with more than one power of $Q_{1}$ or $Q_{2}$ when determining the gauge field
propagator. To get this propagator, we look at the gauge part of the action in momentum space, at leading
order,
\begin{eqnarray}
\label{eq-SA}
S_{A} & = & -\frac{1}{2}\int\frac{d^{4}p}{(2\pi)^{4}} A_{a\mu}\left[p^{2}\eta^{\mu\nu}
-\left(1-\frac{1}{\xi}\right)p^{\mu}p^{\nu} +2Q_{2}u^{\mu}u^{\nu}p^{2}\right. \\
& & \left.{}-4\frac{(\xi-\lambda)}{\xi}Q_{2}u^{\nu}u^{\alpha}p_{\alpha}p^{\mu}
+2Q_{2}u^{\alpha}u^{\beta}p_{\alpha}p_{\beta}\eta^{\mu\nu}\right]A_{a\nu}. \nonumber
\end{eqnarray}
We can read off from the gauge field action $S_{A}$ what the corresponding equation of motion will be.
To determine the propagator, we shall consider a numerator with the possible Lorentz structures
\begin{equation}
\label{eq-Delta}
\Delta_{\mu\delta}=\eta_{\mu\delta}-\zeta\frac{p_{\mu}p_{\delta}}{p^2}+
\beta Q_{2}u_{\mu}u_{\delta}+\gamma Q_{2}u^{\alpha}u_{\delta}\frac{p_{\alpha}p^{\mu}}{p^2}+
\sigma Q_{2}u^{\alpha}u^{\beta}
\frac{p_{\alpha}p_{\beta}}{p^2}\eta_{\mu\delta}
\end{equation}
and insist that it produce a Kronecker $\delta$-function upon contraction with the integrand in (\ref{eq-SA}).
Of course, (\ref{eq-Delta}) is not the most general $p^{\mu}$-dependent matrix that might be used to invert
the bilinear structure in (\ref{eq-SA}). In fact, this $\Delta_{\mu\delta}$ is restricted to have no poles higher
than the ones that appear in the normal gauge boson propagator. The motivation for this is that 
Lorentz violation in the pure gauge sector may be eliminated by a linear
coordinate transformation. (However, only if the SME coefficients satisfy $Q_{1}+2Q_{2}=0$ will the same
transformation that sets $Q_{2}=0$ will also eliminate the apparent Lorentz violation
in the scalar sector, by also setting $Q_{1}=0$.) Therefore, we would like to find a propagator for the gauge
field that resembles the propagator for a Lorentz-invariant non-Abelian gauge field, but expressed in skewed
coordinates --- if this is possible.

The first thing to notice is that $\zeta=(1-\xi)$ has to be satisfied, confirming the usual behavior in the
absence of the Lorentz violation. Moreover, upon contracting and keeping the terms which are first order in
$Q_{2}$, we find a total of twelve terms, which, taken together, should vanish. They are
\begin{eqnarray}
0 & = & \beta u^{\nu}u_{\delta}p^2
+\gamma u_{\alpha}u_{\delta}p^{\alpha}p^{\nu}
+\sigma u^{\alpha}u^{\beta}p_{\alpha}p_{\beta}\delta_{\delta}^{\nu} \\
& & {} -\beta \left(1-\frac{1}{\xi}\right)u_{\alpha}u_{\delta}p^{\alpha}p^{\nu}
-\gamma \left(1-\frac{1}{\xi}\right)u_{\alpha}u_{\delta}p^{\alpha}p^{\nu}
-\sigma \left(1-\frac{1}{\xi}\right)u^{\alpha}u^{\beta}p_{\alpha}p_{\beta}\frac{p_{\delta} p^{\nu}}{p^2}
\nonumber\\
& & {} +2u^{\nu}u_{\delta}p^{2}
-4\frac{\xi-\lambda}{\xi}u^{\alpha}u^{\nu}p_{\alpha}p_{\delta}
+2u^{\alpha}u^{\beta}p_{\alpha}p_{\beta}\delta_{\delta}^{\nu}\nonumber \\
& & {} -2(1-\xi)u^{\alpha}u^{\nu}p_{\alpha}p_{\delta}
+4(1-\xi)\frac{\xi-\lambda}{\xi}u^{\alpha}u^{\nu}p_{\alpha}p_{\delta}\
-2(1-\xi)u^{\alpha}u^{\beta}p_{\alpha}p_{\beta}\frac{p_{\delta}p^{\nu}}{p^2} \nonumber.
\end{eqnarray}
For this to be zero, the set of terms with each different contraction structure of $u^{\mu}$ and $p^{\mu}$ must
vanish. For this to happen, the first and seventh terms together require $\beta=-2$.
Similarly, the third and ninth terms require $\sigma=-2$.
However, with this $\sigma$, we see that we cannot generally cancel the sixth and the twelfth term.

If this complication is temporarily ignored, we can also compute $\gamma=2(1-\xi)$ and
$\lambda=\frac{1+\xi}{2}$, which depend on the gauge parameter $\xi$. The latter expression may appear
particularly problematic, as it seems peculiar that the quantity
$\kappa_{G}^{\mu\nu}$ describing the extent of the Lorentz violation in the ghost loops should depend
explicitly on $\xi$. A further problem may crop up with a 
more general $\kappa^{\mu\nu\alpha\beta}$ than (\ref{eq-kappa}), for which
the propagator $\Delta_{\mu\nu}$ may not be manifestly symmetric in its Lorentz indices $\mu$ and $\nu$.
However, in the Feynman gauge, $\xi=1$, all of these problems are circumvented. Moreover, this fixes the value
$\lambda=1$, so that the Lorentz violation in the kinetic terms for the gauge and ghost fields enter with
the same physical magnitudes.

If we insisted on using a different gauge, then the simple $\Delta_{\mu\delta}$ from (\ref{eq-Delta}) would
not be sufficient to describe the gauge boson propagator. A more complicated structure, with higher poles,
would be required~\cite{ref-cambiaso}.
However, the analysis in Ref.~\cite{ref-cambiaso} is also simplified by the that
fact that it only considers a massive Abelian gauge theory. In a theory
with a vector mass (whether of generalized Proca or Stueckelberg form), there is a consistency condition on
the gauge field; this is a Lorentz-violating generalization of the Lorenz condition
obeyed by the Abelian vector potential in the presence of a Proca mass. This generalized consistency condition
may be used to simplify the action and the propagator for the physical vector boson modes, but this
useful simplifying condition unfortunately becomes trivial when the physical photon mass vanishes.

Fortunately, by using the Feynman gauge --- which is generally the most convenient gauge to use anyway --- all
the complications and caveats are avoided. With the propagators established,
in the next section we shall compute the one-loop radiative corrections to the $\kappa^{\mu\nu\alpha\beta}$
term in the gauge field action, as generated by the coupling the gauge field to virtual scalar matter loops.
To evaluate the relevant correlation functions, we shall use adapted versions of a set of Mathematica
packages~\cite{feyncalc,feynarts,feynrules}. Note, however, that the contributions coming solely from gluon
and ghost propagators (both of which are manifestations of the non-Abelian
gauge sector) have already been
calculated~\cite{ref-collad-3}, although not directly using the gauge fixing prescription described in this
section.

\section{Corrections to the Non-Abelian Aether-Like Term}

\label{sec-scalar-loops}

\begin{figure}[ht]
\begin{centering}
\includegraphics[angle=0,width=14cm]{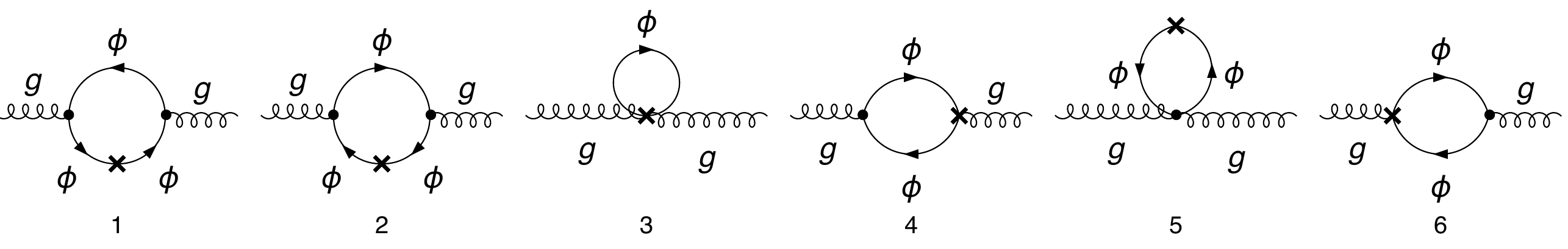}
	\caption{Feynman diagrams for the LV corrections to the gluon field self-energy. Continuous and
	curly lines represent scalar and gluon propagators, respectively. The crossed vertices correspond to
	LV vertex insertions.}
\label{fig01}
\end{centering}
\end{figure}

\begin{figure}[ht]
\begin{centering}
	\includegraphics[angle=0,width=14cm]{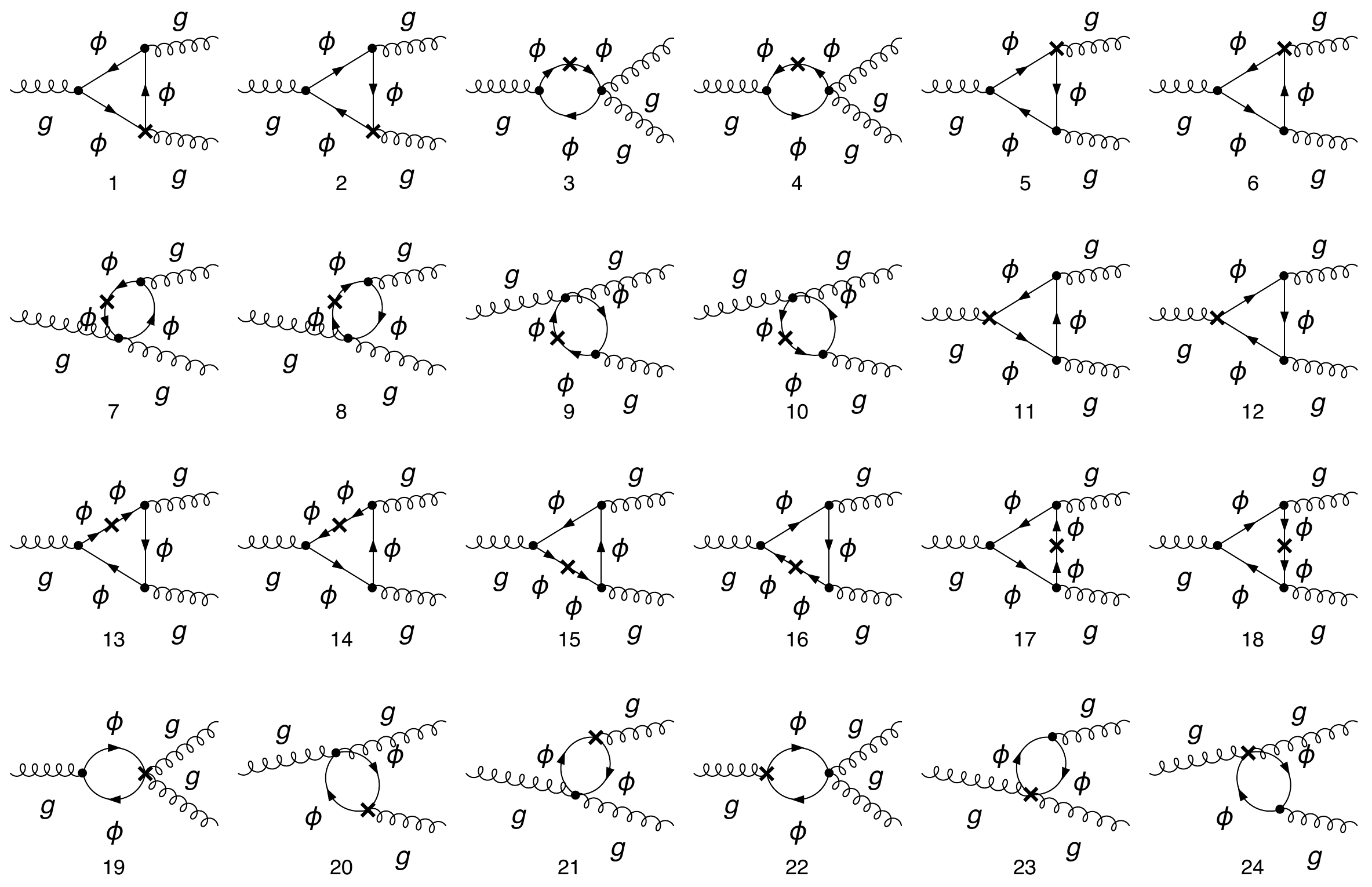}
	\caption{Feynman diagrams for the LV corrections to the three-gluon vertex interaction.}
	\label{fig02}
\end{centering}
\end{figure}

\begin{figure}[ht]
\begin{centering}
	\includegraphics[angle=0,width=12cm]{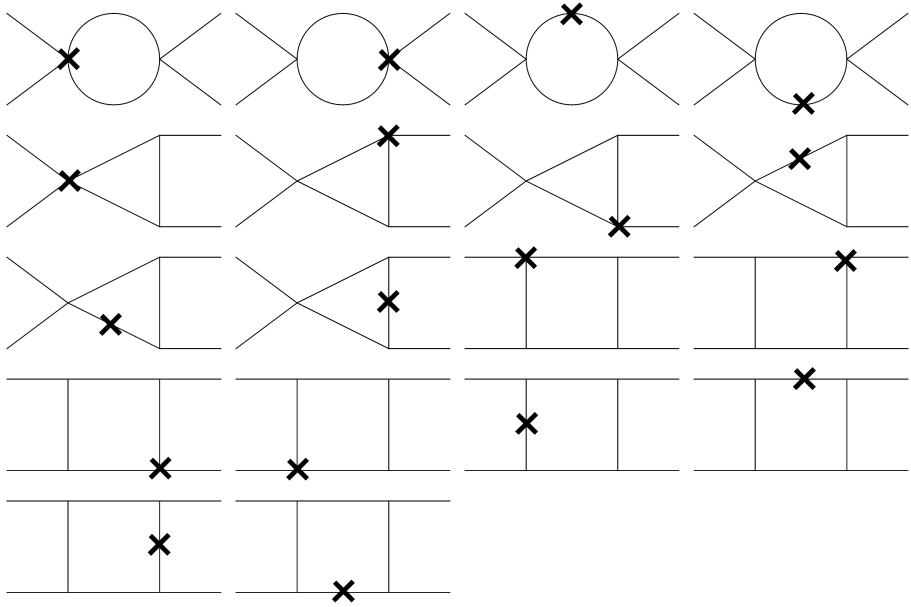}
	\caption{Topologies for the four-point function Feynman diagrams with single LV vertex insertions.
	Both gauge and scalar internal lines are implicitly included.}
	\label{fig03}
\end{centering}
\end{figure}

Let us start with evaluating the matter loop corrections to the non-Abelian aether term
${Q_2}u_{\mu}u_{\nu}F_a^{\mu\alpha}{F_a^{\nu}}_{\alpha}$ appearing in (\ref{eq01}).
The diagrams we must compute are depicted in the figures~\ref{fig01}, \ref{fig02}, and \ref{fig03}. 

First, we calculate  the gluon self-energy given by the Feynman diagrams in figure~\ref{fig01}.
The general structure of the aether-like LV gluon self-energy has the form
\begin{equation}
\label{aether1}
\langle {\rm T}\,A_a^{\mu}(p_1)A_b^{\nu}(p_2) \rangle=\left(-\frac{g^{2}C_{A}Q_{1}}{48 \pi^2 \epsilon}
+\mathrm{finite} \right)\Pi^{\mu\nu}(p_1)\operatorname{tr}(T_{a}T_{b}) (2\pi)^4\delta^{(4)}(p_1+p_2),
\end{equation}
where the representation constant $C_{A}=N$ for the for the adjoint representation of the
$SU(N)$ gauge group, and the Lorentz structure is
\begin{equation}
\Pi^{\mu\nu}(p_1)=\eta^{\mu \nu} (p_1\cdot u)^2-(p_1^{\mu } u^{\nu }
+p_1^{\nu} u^{\mu}) (p_1\cdot u)+p_1^2 u^{\mu } u^{\nu};
\end{equation}
here we have used the conservation law $p_2=-p_1$ for the external momentum.
In this and the following calculations we define, as usual, the dimensional
extension $\epsilon = \frac{D-4}{2}$.

The matter corrections to the aether-like LV three-gluon vertex, figure~\ref{fig02}, can similarly
be cast as
\begin{eqnarray}
\label{aether2}
\langle {\rm T}\,A_a^{\alpha}(p_1)A_b^{\mu}(p_2)A_c^{\nu}(p_3) \rangle & = &
\left(-\frac{g^{2}C_{A}Q_{1}}{48 \pi^2 \epsilon} +\mathrm{finite}\right)\Pi^{\alpha\mu\nu}(p_1,p_2,p_3) \\
& & \times\operatorname{tr}([T_a,T_b]T_c)(2\pi)^4\delta^{(4)}(p_1+p_2+p_3), \nonumber
\end{eqnarray}
where
\begin{eqnarray}
\Pi^{\alpha\mu\nu}(p_1,p_2,p_3)&=& (p_1-p_3)^\alpha u^\mu u^\nu +(p_3-p_2)^\nu u^\alpha u^\mu
+(p_2-p_1)^\mu u^\nu u^\alpha \\
& & {} +u^\alpha \eta^{\mu\nu}(p_1-p_3)\cdot u +u^\nu \eta^{\mu\alpha}(p_3-p_2)\cdot u
+u^\mu \eta^{\alpha\nu}(p_2-p_1)\cdot u. \nonumber
\end{eqnarray}
Naturally, the symbol $[T_{a},T_{b}]$ stands for the commutators between the normalized group generators,
$[T_{a},T_{b}]=if_{abc}T_{c}$.

Finally, matter loop corrections to the aether-like LV four-gluon vertex, shown in figure~\ref{fig03},
are given by
\begin{eqnarray}
\label{aether3}
\langle {\rm T}\,A_a^{\alpha}(p_1)A_b^{\beta}(p_2)A_c^{\mu}(p_3)A_d^{\nu}(p_4) \rangle &=&
\left(-\frac{g^{2}C_{A}Q_{1}}{48 \pi^2 \epsilon}+\mathrm{finite}\right)
\Pi^{\alpha\beta\mu\nu}(p_1,p_2,p_3,p_4) 
\operatorname{tr}\big([T^a,T^d][T^b,T^c] \nonumber\\
& & {} +[T^a,T^c][T^b,T^d]\big)(2\pi)^4\delta^{(4)}(p_1+p_2+p_3+p_4),
\end{eqnarray}
where in this case
\begin{eqnarray}
\Pi^{\alpha\beta\mu\nu}(p_1,p_2,p_3,p_4) & = & 
u^\alpha u^\beta \eta^{\mu\nu} +u^\mu u^\nu \eta^{\alpha\beta}
+u^\alpha u^\nu \eta^{\beta\mu} \nonumber\\
& & {}+u^\beta u^\mu \eta^{\alpha\nu}-2 (u^\alpha u^\mu \eta^{\beta\nu} +u^\beta u^\mu \eta^{\beta\nu}).
\end{eqnarray}

Using the Lie algebra of the generators $T_{a}$ of the $SU(N)$ gauge group and combining
the results (\ref{aether1}), (\ref{aether2}), and (\ref{aether3}),
we obtain the following effective Lagrange density
\begin{eqnarray}
\mathcal{L}_{{\rm eff}}=\operatorname{tr}\left[\left(Z_{Q_2}Q_2-\frac{g^{2}C_{A}Q_{1}}{96 \pi^2 \epsilon}
+\mathrm{finite}\right)u_{\mu}u_{\nu}F^{\mu\alpha}{F^{\nu}}_{\alpha}\right],
\end{eqnarray}
where $Z_{Q_2}Q_2=(Q_2+\delta_{Q_2})$, with $\delta_{Q_2}$ being the appropriate counterterm,
which in the minimal subtraction (MS) renormalization scheme is evidently given by
\begin{eqnarray}
\delta_{Q_2}=\frac{g^{2}C_{A}Q_{1}}{96 \pi^2 \epsilon}.
\end{eqnarray}
This is one of the pieces needed to determine how the RG behavior of the theory is affected by the
charged scalars.
In the next section we compute the scalar self-energy contribution, as well as the gluon-matter contribution,
in order to complete the computation the $\beta$-functions for the model.

\section{Scalar Self-Energy and the Gluon-Matter Vertex}

\label{sec-scalar}

\begin{figure}[ht]
\begin{centering}
	\includegraphics[angle=0,width=12cm]{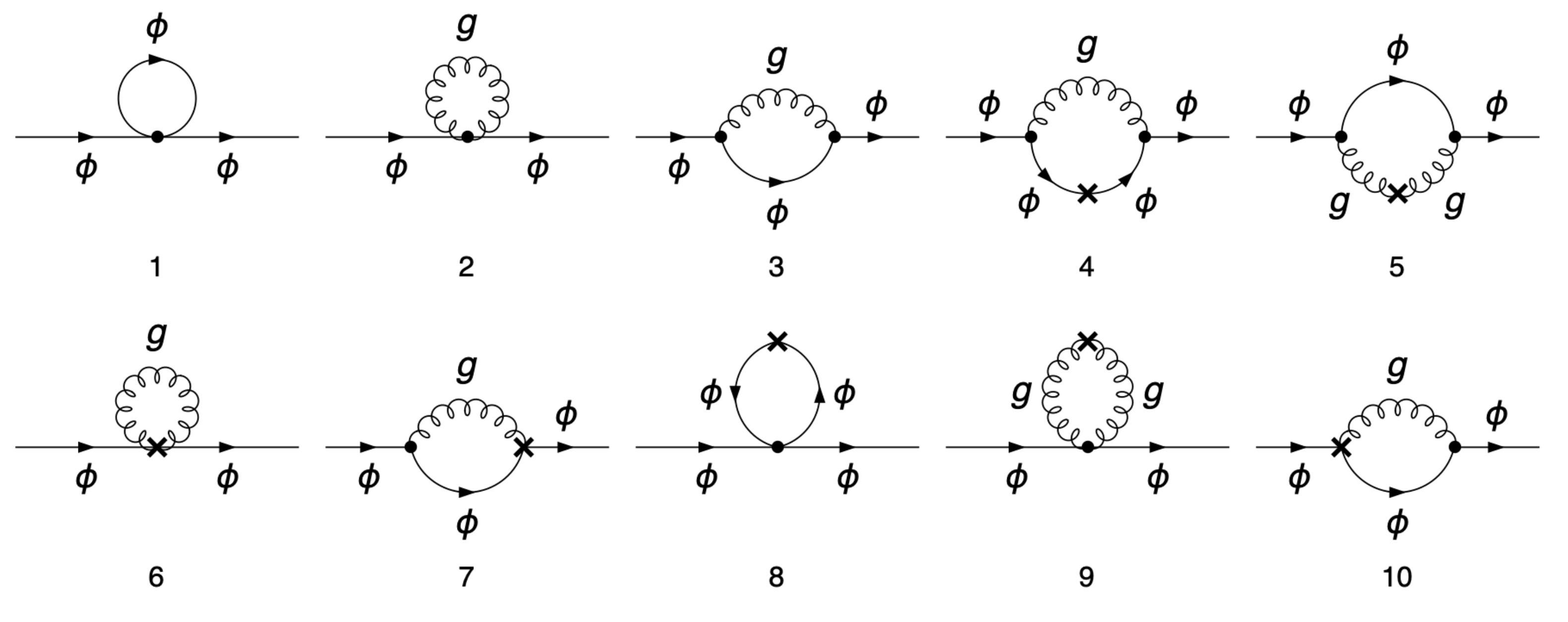}
	\caption{Feynman diagrams for the scalar field self-energy.}
	\label{fig04}
\end{centering}
\end{figure}

Let us now compute the one-loop scalar self-energy (figure~\ref{fig04}),
$\Sigma(p)=\langle {\rm T}\,\phi_a^i(p)\phi_b^{j*}(-p)\rangle$.
The expression correponding to figure~\ref{fig04}'s diagram~1 (or ``diagram~\ref{fig04}--1'') is given by
\begin{eqnarray}\label{eqfig01-1}
\Sigma_1(p) & = &-i\!\int\!\!\frac{d^{4}k}{(2\pi)^{4}}
\frac{2\lambda N_{s}}{\left(k^{2}-m^{2}\right)}\delta_{ij}\delta_{ab} \\
& = & \frac{\lambda N_{s}}{4\pi^2}A_0(m^2)\delta_{ij}\delta_{ab},
\end{eqnarray}
where $A_0(m^2)$ is one of the Passarino-Veltman basis integrals (see appendix~\ref{appendixPV}),
and $N_{s}$ is the number of scalar fields [which must be a multiple of $(N^{2}-1)$ for $SU(N)$ adjoint scalars].

The expression for the tadpole diagram \ref{fig04}--2 is proportional to $A_0(0)$ --- that is, vanishing
(up to irrelevant, regulator-dependent finite terms).
However, the corresponding expression for diagram~\ref{fig04}--3 is
\begin{eqnarray}
\label{eqfig01-3}
\Sigma_3(p) & = & i\!\int\!\!\frac{d^4k}{(2\pi)^4}
\frac{g^2\eta_{\mu\nu}(k+p)^\mu (k+p)^\nu f^{acd}f^{bcd}}{\left(k^2-m^2\right)\left(k-p\right)^2}\delta_{ij} \\
& = & \frac{g^{2}C_{A}}{16\pi^2}\left[A_0(m^2)-2(p^{2}+m^{2})B_0(p^2,0,m^2)\right]\delta_{ij}\delta_{ab},
 \end{eqnarray}
where $B_0(p^2,0,m^2)$ is another integral from the Passarino-Veltman basis.

The expression for the first diagram with a LV vertex insertion, diagram~\ref{fig04}--4, is given by
\begin{eqnarray}
\label{eqfig01-4}
\Sigma_4(p) &=&-i\!\int\!\!\frac{d^4k}{(2\pi)^4} \frac{g^2Q_{1}(k+p)^2(k\cdot u)^2f^{acd}f^{bcd}}
{\left(k^2-m^2\right)^2\left(k-p\right)^2}\delta_{ij} \\
& = & -\frac{g^{2}Q_{1}C_{A}}{24\pi^{2}}\frac{(p\cdot u)^{2}}{p^{4}}
\Big[2(p^2+m^2)A_0(m^2)  \\
& & {} -(2p^4+4m^2p^2+3m^4)B_0(p^2,0,m^2) +(p^2+m^2)^2B_0(0,m^2,m^2) \nonumber\\
& & {} -(p^6+2m^2p^4+2m^4p^2+m^6)C_0(0,p^2,p^2,m^2,m^2,0)\Big]\delta_{ij}\delta_{ab}. \nonumber
\end{eqnarray}
Here we have encountered the last of the Passarino-Veltman basis integrals that we need; like the others,
$C_0(0,p^2,p^2,m^2,m^2,0)$ is given in appendix~\ref{appendixPV}.

Continuing, the expression for diagram~\ref{fig04}--5 is
\begin{eqnarray}
\label{eqfig01-5}
\Sigma_5(p) & = & -i\!\int\!\!\frac{d^{4}k}{(2\pi)^4} \frac{g^{2}Q_{2}[p^2 (k\cdot u)^2+k^2(p\cdot u)^2
-2(k\cdot p)(k\cdot u)(p\cdot u)]f^{acd}f^{bcd}}{2\left(k^2\right)^2\left[(k-p)^2-m^2\right]}
\delta_{ij}\,\,\,\, \\
& = &  -\frac{g^{2}Q_{2}C_{A}}{192\pi^{2}}\frac{(p\cdot u)^{2}}{p^{2}}
\Big[ A_0(m^2)-2(p^2+m^2)B_0(p^2,0,m^2) \\
& & {} -(p^2-m^2)B_0(0,0,0)+(p^2-m^2)^2C_0(0,p^2,p^2,0,0,m^2)\Big]\delta_{ij}\delta_{ab}. \nonumber
\end{eqnarray}

Diagram~\ref{fig04}--6 is once again a tadpole proportional to $A_0(0)=0$. The following diagram,
\ref{fig04}--7, can be expressed as
\begin{eqnarray}
\label{eqfig01-7}
\Sigma_7(p) & = & i\!\int\!\!\frac{d^4k}{(2\pi)^4}
\frac{g^{2}Q_{1}[(k+p)\cdot u]^2f^{acd}f^{bcd}}{\left(k^2-m^2\right)^2(k-p)^2}\delta_{ij} \\
& = & \frac{g^{2}Q_{2}C_{A}}{48\pi^{2}}\frac{(p\cdot u)^{2}}{p^{4}}\Big[ (4p^2+m^2)A_0(m^2) \\
& & {} -(7p^4+4m^2p^2+m^4)B_0(p^2,0,m^2)\Big]\delta_{ij}\delta_{ab}. \nonumber
 \end{eqnarray}

The expression corresponding to diagram~\ref{fig04}--8 is proportional to the trace of the LV tensor
$c^{\mu\nu}$ --- that is to $u^{2}=0$. Since diagram~\ref{fig04}--9 is tadpole, it can also be expressed as
\begin{eqnarray}\label{eqfig01-9}
\Sigma_9(p) & = & i\!\int\!\!\frac{d^4k}{(2\pi)^4}
\frac{g^{2}Q_{2}(k\cdot u)^2 f^{acd}f^{bcd}}{4(k^2)^2}\delta_{ij} \propto A_0(0)=0.
\end{eqnarray}
Finally, the expression for diagram~\ref{fig04}--10 is
\begin{eqnarray}
\label{eqfig01-10}
\Sigma_{10}(p) & = & i\!\int\!\!\frac{d^4k}{(2\pi)^4}
\frac{g^{2}Q_{1}[(k-p)\cdot u]^2 f^{acd}f^{bcd}}{(k^2-m^2)(k+p)^2}\delta_{ij}=\Sigma_{7}(p) .
 \end{eqnarray}

Adding all the above expressions, along with the counterterm diagrams, we find
\begin{eqnarray}
\label{eqfig01-UV}
\Sigma(p) & = & \Bigg[\frac{4\lambda N_s m^2-g^{2}C_{A}(2p^2+m^2)}{16\pi^2\epsilon}
-\frac{g^{2}(Q_1+Q_2)C_{A}(p\cdot u)^2}{4\pi^2\epsilon} \\
& & (\delta_2 p^2-\delta_{m^2} m^2)+\delta_{Q_1}(p\cdot u)^2+\mathrm{finite}\Bigg]\delta_{ij}\delta_{ab},
\nonumber
\end{eqnarray}
where $\delta_{2}$, $\delta_{m^{2}},$ and $\delta_{Q_{1}}$ are the counterterms in question, with the
forms:
$\delta_2=Z_2-1$, $\delta_{m^2}=(m_0^2Z_{m^2}-m^2)/m^2$, and $\delta_{Q_1}=Q_1(Z_{Q_1}-1)$. Here, $Z_2$,
$Z_{m^2}$, and $Z_{Q_1}$ are the corresponding renormalization constants (in particular, $Z_2$ is the
renormalization constant for the scalar field redefinition $\phi_a \rightarrow \sqrt{Z_2}\phi_a$), and
$m_{0}^{2}$ is the bare scalar mass squared.
Imposing finiteness of (\ref{eqfig01-UV}) through the MS scheme, we find
\begin{subequations}
\begin{eqnarray}
\delta_2 & = & \frac{g^{2}C_{A}}{8 \pi^2 \epsilon},\label{ct01}\\
\delta_{m^2} & = & \frac{\left(\lambda N_{s} -g^{2}C_{A}\right)}{16 \pi^2 \epsilon}, \label{ct02}\\
\delta_{Q_1} & = & \frac{g^{2}(Q_1+Q_2)C_{A}}{4 \pi^2 \epsilon}. \label{ct03}
\end{eqnarray}
\end{subequations}

\begin{figure}[ht]
\begin{centering}
	\includegraphics[angle=0,width=12cm]{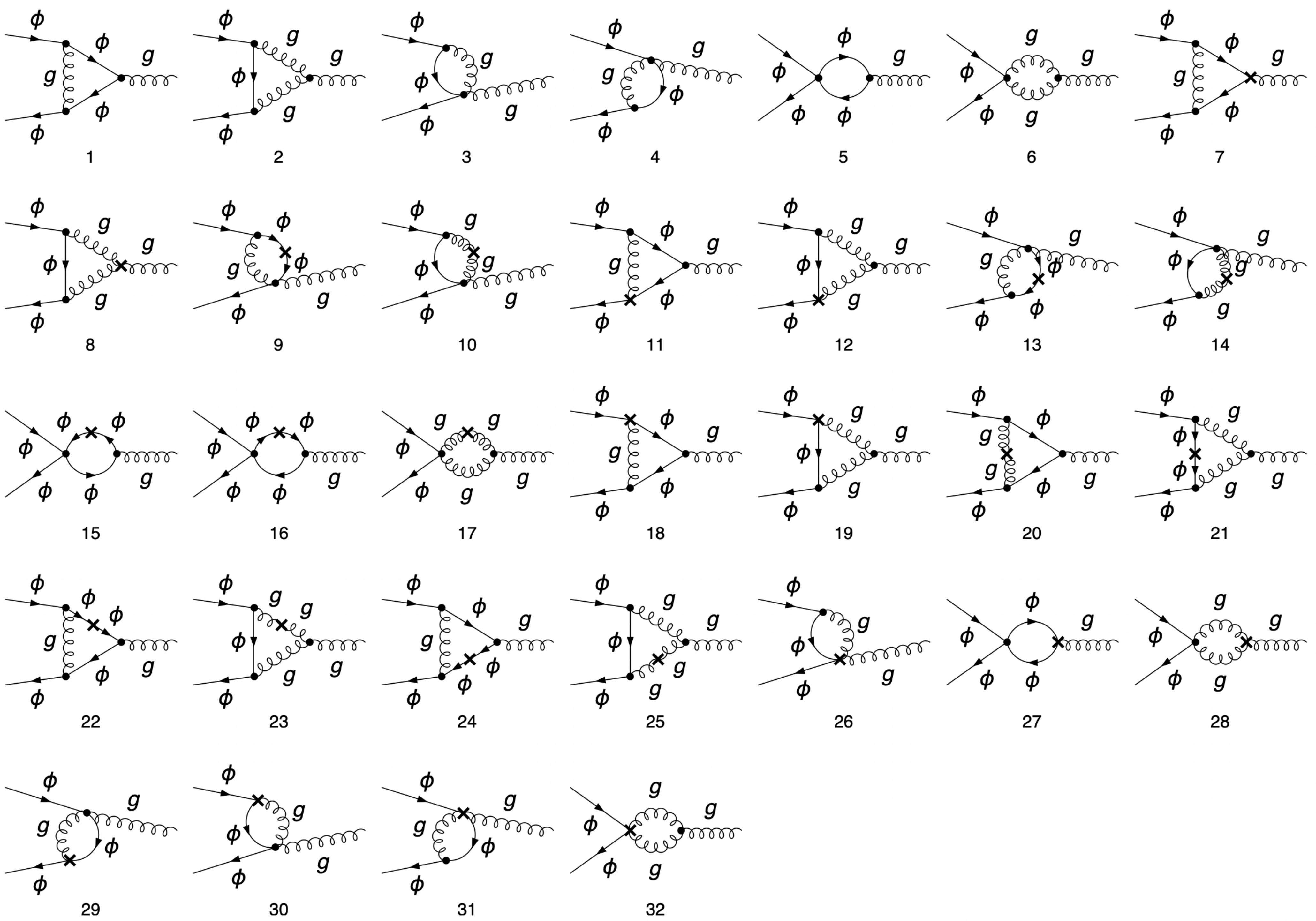}
	\caption{Feynman diagrams for the gluon-matter vertex.}
	\label{fig05}
\end{centering}
\end{figure}

For completeness, we shall also evaluate the radiative corrections to the gluon-matter-current interaction
vertex. The Feynman diagrams of this process are depicted in figure~\ref{fig05}. The general form taken by the
vertex function is
\begin{eqnarray}
\Gamma=i\!\int\!\!\frac{d^4p_1}{(2\pi)^4}\frac{d^4p_2}{(2\pi)^4}\frac{d^4p_3}{(2\pi)^4}
f_{abc}\phi_a(p_1)\phi^*_b(p_2)\Gamma_{\mu} A^\mu_c(p_3)(2\pi)^4\delta(p_1+p_2+p_3).
\end{eqnarray}
The corresponding ultraviolet-divergent expression may be written as
\begin{eqnarray}\label{vertex1}
\Gamma^{\mu}(p_1,p_2,p_3) & = &
\frac{g^{3}C_{A}(p_2-p_1)^{\mu }}{16 \pi^2 \epsilon}+\frac{3g^{3}(Q_1+Q_2)C_{A}[(p_2-p_1)\cdot u]u^{\mu}}
{16 \pi^2 \epsilon} \\
& & {} +\delta_1 g\left(p_2-p_1\right)^{\mu} -\tilde{\delta}_{Q_1}[(p_2-p_1)\cdot u]u^{\mu},
\nonumber
\end{eqnarray}
where $\delta_1$ and $\tilde{\delta}_{Q_1}$ are the appropriate counterterms, and we have
once again used momentum conservation to simplify the final expression. Setting (\ref{vertex1})
to be vanishing, we find the counterterm values
\begin{subequations}
\begin{eqnarray}
\delta_1 & = & \frac{g_{0}Z_{2}\sqrt{Z_{3}}-g}{g}=-\frac{g^{2}C_{A}}{16 \pi^2 \epsilon}, \label{ct06}\\
\tilde{\delta}_{Q_1} & = & \frac{g_{0}(Q_{1})_{0}Z_{2}\sqrt{Z_{3}}-Q_{1}}{Q_{1}}= \frac{3g^{2}(Q_1+Q_2)C_{A}}
{16 \pi^2 \epsilon}, \label{ct07}
\end{eqnarray}
\end{subequations}
where $Z_3$ is the renormalization constant for the gluon field redefinition
$A_\mu\rightarrow\sqrt{Z_{3}}A_\mu$.
The counterterm $\delta_3=Z_3-1$ is computed in appendix~\ref{appendixGSE}.

It is important to mention that
the renormalization of the $\lambda\phi^4$ operator is just like in the ordinary theory, since the LV
corrections are necessarily proportional to $u^2=0$. Any LV corrections to the scalar
vertex must involve contractions of $u^{\mu}$ with derivatives of the field, making them manifestly
finite by power counting.

\section{$\beta$-Functions}

We now proceed to compute the scalar-matter coupling corrections to the one-loop $\beta$-functions
for the of the Lorentz violation paramters
$Q_{1}$ and $Q_{2}$. In the previous sections we have found the one-loop
counterterms for the scalar and gluon fields and the gluon-matter vertex in our model. In order to evaluate
the relevant $Z$ factors to compute the $\beta$-functions of $Q_{1}$ and $Q_{2}$, we expand each
renormalization constant $Z_i$ as power series in the coupling constants, which can be determined
order by order using perturbation theory,
\begin{eqnarray}
Z_i=1+\delta^{(1)}_i+\delta^{(2)}_i+\cdots.
\end{eqnarray}
The relation between the bare and renormalized coupling constants may be cast as
\begin{subequations}
\label{Zcouplings}
\begin{eqnarray}
Z_{Q_1}Q_1 & = & \mu^{-2\epsilon}(Q_{1})_{0}Z_2=Q_1+\delta_{Q_1},\\
Z_{Q_2}Q_2 & = & \mu^{-2\epsilon}(Q_{2})_{0}Z_3=Q_2+\delta_{Q_2},
\end{eqnarray}
\end{subequations}
where, as noted above, $Z_2$ ($\phi_a\rightarrow \sqrt{Z_{2}}\phi_a$) and $Z_3$
($A_a^\mu\rightarrow \sqrt{Z_{3}}A_a^\mu$) are the field strength renormalization constants for
the scalar and gauge boson fields, respectively.

Taking the scaling relations (\ref{Zcouplings}) between the  bare and renormalized couplings,
we can compute $\beta_{Q_{1}}$ to be
\begin{equation}
\beta_{Q_{1}}= \lim_{\epsilon\rightarrow 0}
\left[-2\epsilon {Q_1}\left(1+\delta_2-\frac{\delta_{Q_1}}{Q_1}\right)\right] 
 =\frac{g^{2}(Q_1+2Q_2)C_{A}}{4 \pi^{2}}. \label{betaq1}
\end{equation}
When $Q_{1}+2Q_{2}=0$, there is no RG flow for this SME parameter.
Similarly, the scalar matter-loop fluctuations result in the following corrections to the
$\beta_{Q_{2}}$ function:
\begin{equation}
\beta_{Q_{2}}= \lim_{\epsilon\rightarrow 0}\left[-2\epsilon {Q_2}
\left(1+\delta_3-\frac{\delta_{Q_2}}{Q_2}\right)\right]=\frac{g^{2}N_{s}
(Q_1+2Q_2)}{48 \pi^2}+\cdots, \label{betaq2}
\end{equation}
\noindent where the dots mean that we have included only the matter-loop corrections to $\beta_{Q_{2}}$.

\section{Final Remarks}

\label{summary}
 
Renormalization is an essential part of the work to understand a quantum field theory.
The regularization and renormalization processes allow us to verify the consistency of a specific
field-theoretic model at the quantum level and to understand its underlying behavior at high energy
scales. Even in theories, such as the SME, in which some of the core principles of relativistic field
theory may  have been discarded, it is still crucial to understand the theory's renormalization
properties --- either by generalizing and reformulating formal
renormalization theorems or by explicit perturbative calculations. It is,
in fact, one of the greatest strengths of the SME approach that, since it is an effective field theory,
it is amenable to calculation of radiative corrections in essentially just the same way as the usual SM.

This paper has addressed the case where Lorentz symmetry is broken, but non-Abelian gauge symmetries are
still preserved. More specifically, we formulated the LV extension of a non-Abelian gauge theory coupled
to a scalar matter. We calculated the contributions to scalar and gluon field strength renormalization,
as well as the to vector-scalar interaction vertex, at the lowest orders in the gauge couplings and the
Lorentz violation parameters. Using these results, we computed the one-loop $\beta$-functions for the model.
From
the forms of $\beta_{Q_{1}}$ and $\beta_{Q_{2}}$, it is evident that there is no running of these coupling
constants when there is a particular relationship between the SME coefficients in the gauge and matter sectors,
specifically when $Q_1+2Q_2=0$. When this relation is satisfied, the LV couplings are scale independent at
this order; this corresponds to a theory in which the Lorentz violation may be eliminated from both sectors
by a transformation to skewed four-dimensional Cartesian coordinates.
The inclusion of quantum corrections should not change the fact that the theory is
actually Lorentz invariant, but merely expressed in oblique coordinates.

The theory we have considered is, more generally, part of the non-Abelian gauge and scalar sectors of the
LV SME. As we have already noted, there is still a significant amount of further work to be done to understand
these types of theories, with couplings between of non-Abelian gauge fields and charged scalar matter --- in
particular, concerning the explicit calculation of radiative corrections. The present contribution gives a
first step in this direction, but there are important topics that have still not been touched, such as
the effects of spontaneous gauge symmetry breaking. Since the electroweak sector of the SM includes a multiplet
of scalar Higgs fields that break $SU(2)_{L}$ gauge invariance, this is an important area in which further
research will need to be undertaken in the future.

\section*{Acknowledgments}

The authors are grateful to J. R. Nascimento for important discussions.
The work of A. Yu.\ P. has been partially supported by the CNPq project No. 301562/2019-9.

\appendix

\section{Passarino-Veltman Basis Integrals}
\label{appendixPV}

Many of the integrals appearing in section~\ref{sec-scalar} were reduced to linear combinations of 
the following standard integrals, which are components of the Passarino-Veltman basis:
\begin{eqnarray}
A_0(m^2) &=& \int\!\!{d^{D}k}\,\frac{1}{k^2-m^2}=\frac{m^2}{\epsilon}+\mathrm{finite};\label{PV1}\\
B_0(p^2,m_1^2,m_2^2) &=& \int\!\!{d^{D}k}\,\frac{1}{(k^2-m_1^2)((k+p)^2-m_2^2)}=\frac{1}{\epsilon}+\mathrm{finite};\label{PV2}\\
C_0(p_1^2,(p_1-p_2)^2,p_2^2,m_1^2,m_2^2,m_3^2) &=& \int\!\!{d^{D}k}\,\frac{1}{(k^2-m_1^2)((k+p_1)^2-m_2^2)((k+p_2)^2-m_3^2)}\nonumber\\
&=&\mathrm{finite} \label{PV3}.
\end{eqnarray}

\section{Gluon self-energy}\label{appendixGSE}

\begin{figure}[ht]
\begin{centering}
	\includegraphics[angle=0,width=12cm]{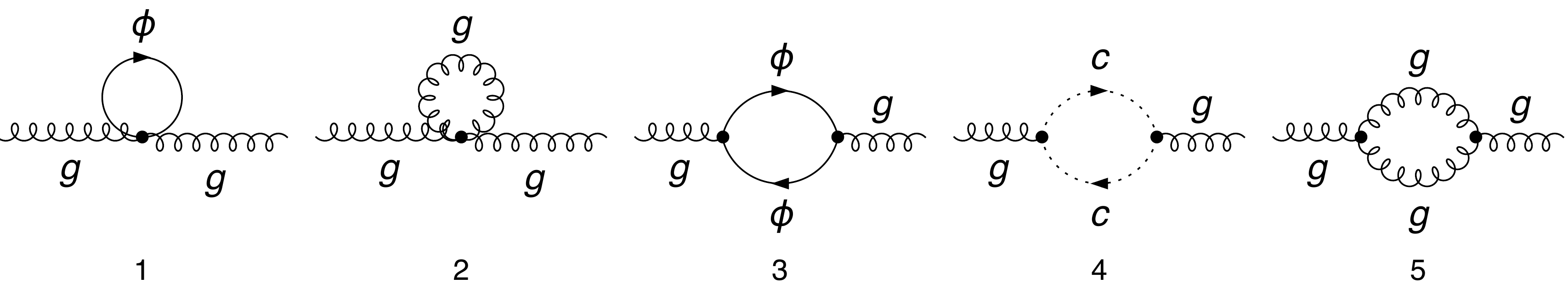}
	\caption{Feynman diagrams for the ordinary gluon self-energy. Dashed lines represent Faddeev-Popov
	ghost propagators.}
	\label{fig06}
\end{centering}
\end{figure}

The Feynman diagrams for the ordinary one-loop gluon self-energy are depicted in figure~\ref{fig06}.
The general structure of the gluon two-point function has the form
\begin{equation}
\Gamma_A=\int\!\!\frac{d^4p}{(2\pi)^4}A_a^{\mu}(p)\Pi_{\mu\nu}(p)A_a^{\nu}(-p).
\end{equation}
The expressions corresponding to the individual diagrams shown in figure~\ref{fig06} are
\begin{eqnarray}
\Pi_1^{\mu\nu}(p) & = & i\!\int\!\!\frac{d^4k}{(2\pi)^4}\frac{2g^{2}N_{s}\eta^{\mu\nu}}{(k^2-m^2)}
= -\frac{g^{2}N_{s}}{8\pi^2}\eta^{\mu\nu}A_0(m^2), \\
\Pi_3^{\mu\nu}(p) & = & -i\!\int\!\!\frac{d^4k}{(2\pi)^4}\frac{g^{2}N_{s}(2k-p)^\mu(2k-p)^\nu}
{(k^2-m^2)((k-p)^2-m^2)} \\
& = & \frac{g^{2}N_{s}}{48\pi^{2}}\frac{1}{p^{2}}\left[(p^2\eta^{\mu\nu}+2p^{\mu}p^{\nu})A_0(m^2)
+(4m^2-p^2)(p^2\eta^{\mu\nu}-p^\mu p^\nu)B_0(p^2,m^2,m^2)\right], \nonumber\\
\Pi_4^{\mu\nu}(p) & = & i\!\int\!\!\frac{d^4k}{(2\pi)^4}\frac{g^{2}C_{A}k^\mu (k-p)^\nu}{k^2(k-p)^2}
=\frac{g^{2}C_{A}}{192\pi^2}(p^2\eta^{\mu\nu}+2p^\mu p^\nu)B_0(p^2,0,0), \\
\Pi_5^{\mu\nu}(p) & = & -i\!\int\!\!\frac{d^4k}{(2\pi)^4}\frac{g^{2}C_{A}
\left[\eta^{\mu\nu}\left(5p^2+2k^2-2(k\cdot p)\right)-p^{\mu}
\left(5k^\nu+2p^\nu\right)+5k^\mu\left(2k^\nu-p^\nu\right)\right]}{2k^2(k-p)^2}\nonumber\\
&=& \frac{g^{2}C_{A}}{192\pi^2}(p^2\eta^{\mu\nu}+2p^\mu p^\nu)B_0(p^2,0,0),
\end{eqnarray}
while diagram~\ref{fig06}--2, being a tadpole, is proportional to $A_0(0)=0$.

Adding these contributions, substituting the integrals (see again appendix~\ref{appendixPV}),
and keeping only the ultraviolet-divergent terms, we have
\begin{equation}
\label{lv-gluon-structure}
\Pi^{\mu\nu}(p)= \left(p^2\eta^{\mu\nu}-p^\mu p^\nu\right)\left[\frac{g^{2}(5C_A-N_s)}{12\pi^{2}\epsilon}
-\delta_3\right].
\end{equation}
Imposing finiteness, we immediately find 
\begin{equation}
\delta_3=\frac{g^{2}(5C_{A}-N_{s})}{12 \pi ^2 \epsilon}.
\end{equation}
Notice that if we have only one octet of QCD scalar fields, we will have $N_{s}=N^{2}-1=8$
and $C_{A}=N=3$.

\end{document}